\documentclass[preprint,review]{elsarticle}

\usepackage{graphicx}
\usepackage{amssymb}
\usepackage{natbib}
\usepackage{lineno}
\usepackage{multirow}
\usepackage{hhline}
\usepackage{subfigure}
\usepackage{rotating}

\usepackage{fancyhdr}
\fancypagestyle{pprintTitle}{%
  \lhead{}\lfoot{
    DOI: https://doi.org/10.1016/j.radphyschem.2018.01.023 \\
 \ \\
    \copyright \ 2018. This manuscript version is made available under the CC-BY-NC-ND 4.0 license http://creativecommons.org/licenses/by-nc-nd/4.0/}\cfoot{}\rfoot{}
  
}
\journal{Radiation Physics and Chemistry}

\begin{document}
\newcommand{\Angst}{$\mathring{\mathrm{A}}$}

\begin{frontmatter}

\title{Measurements and Monte-Carlo simulations of the particle self-shielding effect of B$_{4}$C grains in neutron shielding concrete}

\author[a,b]{D. D. DiJulio\corref{cor1}}
\ead{Douglas.DiJulio@esss.se}

\author[a,c]{C. P. Cooper-Jensen}

\author[a,d]{I. Llamas-Jansa}

\author[d]{S. Kazi}

\author[a,c]{P. M. Bentley}

\cortext[cor1]{Corresponding Author}

\address[a]{European Spallation Source ERIC, P.O. Box 176, SE-221 00 Lund, Sweden}
\address[b]{Division of Nuclear Physics, Lund University, SE-221 00 Lund, Sweden} 
\address[c]{Department of Physics and Astronomy, Uppsala University, SE-751 20 Uppsala, Sweden}
\address[d]{Institute for Energy Technology, P.O. Box 40,  2027 Kjeller, Norway}

\begin{abstract}
  A combined measurement and Monte-Carlo simulation study was carried out in order to characterize the particle self-shielding effect of B$_{4}$C grains
  in neutron shielding concrete. Several batches of a specialized neutron shielding concrete, with varying B$_{4}$C grain sizes, were exposed to
  a 2 {\AA} neutron beam at the R2D2 test beamline at the Institute for Energy Technology located in Kjeller, Norway.
  The direct and scattered neutrons were detected with a neutron detector placed behind the concrete blocks and the results were
  compared to Geant4 simulations. The particle self-shielding effect was included in the Geant4 simulations by calculating effective neutron cross-sections
  during the Monte-Carlo simulation process. It is shown that this method well reproduces the measured results. Our results show that shielding calculations
  for low-energy neutrons using such materials would lead to an underestimate of the shielding required for a certain design scenario if the particle self-shielding effect is not included in the calculations. 
\end{abstract}

%----------------------------------------------------------------------
% Manuscript keywords
%
% Please give two or three keywords in the form: keyword \sep keyword
% e.g. NMR \sep superconductivity
%
% NB The syntax is different from the abstract document class


\hyphenation{}

\end{frontmatter}

%\linenumbers

\section{Introduction}
The use of concrete as a neutron shielding material is common practice at neutron research facilities, nuclear reactors, and hadron therapy treatment facilities. The choice of using concrete is a trade-off
between the shielding characteristics, cost, and other engineering design requirements, such as stability and physical space for example. A common approach
to enhance the shielding characteristics of concrete for low-energy neutrons is to add boron containing compounds to the mixture \cite{Oakridge1970}.
This enhancement is due to the large low-energy neutron absorption cross-section for $^{10}$B and additionally to the low-energy of the emitted secondary photon radiation
after the absorption process \cite{PGNAA}. The latter property helps to limit the dose rate behind a shield from photons produced in low-energy neutron absorption processes in the concrete material itself. One example of such a concrete is the outer layers of the target monolith at the Swiss Spallation Neutron Source \cite{SINQ,Wagner2001}. \\
\indent For energies up to 1 MeV, it has been suggested
in addition to add polyethylene (PE) to improve the neutron slowing down properties of the concrete \cite{Park2014}. Such a concrete can be
especially effective at spallation neutron sources, such as the European Spallation Source (ESS) \cite{ESS} currently under construction in Lund, Sweden,
where a significant number of high-energy neutrons, keV and above, will escape the target into the bulk shielding of the facility \cite{Bauer2001,Koprivnikar2002}. For this reason, we have developed a new specialized neutron shielding concrete
based on the addition of B$_{4}$C grains and PE beads and studied the performance of the concrete in the MeV energy range
in a previous publication \cite{DiJulio2017}. This concrete is referred to as PE-B4C-concrete below. 
In the current work, we report on the low-energy neutron transport properties of the specialized concrete. The results have relevance for any type of shielding material containing a low amount of small neutron
absorbing grains. \\
\indent The effectiveness of a concrete containing a small amount of B$_{4}$C grains depends not only on the weight fraction of B$_{4}$C added to the mixture but also on
the size of the grains. This effect is referred to as the particle self-shielding effect, which can lead to a reduced performance of the additive in the concrete. 
If the diameter of a grain is large enough, the interior region of the grain will be shielded from neutrons by the outer layers of the grain. Thus a fraction
of the boron which was added to the concrete is rendered in-effective. Early studies of the reduced performance of absorbing grains due to this
effect were carried out for neutron transmission in Boral \cite{Oakridge1960,Sweden1960} and in samples containing Al and B$_{4}$C spherical particles \cite{Doub1961}.
Later, an analytical model \cite{Levermore1986} was used to show the limitations of assuming a homogeneous sample for the transmission of neutrons passing through a mixture of sulfur and tungsten grains \cite{Becker2014}.
More recently, HDPE loaded with small B$_{4}$C grains was investigated in a combined simulation and experimental study and it was noted that smaller grain sizes led to improved neutron shielding properties \cite{Soltani2016}. \\
\indent The design of neutron shielding components is frequently carried out using Monte-Carlo simulations. Typically, the material of a shield is assumed to be composed
of a homogeneous random distribution of the elements within the shield and the particle-self shielding effect is not included. If this effect is completely neglected, the shielding calculations would lead to incorrect
predictions of the required thicknesses of the materials needed for the design scenario. A possible way to include this effect is to adapt an effective density for the absorbing additive. However, as the particle-self shielding
effect is energy dependent, such a method would have limited applicability especially over a broad energy spectrum. Alternatively, a study presented by T. Yamamoto \cite{Yamamoto2014} proposed an effective cross-section calculation method for
this application where effective cross-sections of the components in the material were calculated during a Monte-Carlo simulation. The effective cross-sections took into account the particle self-shielding effect of the absorber grains. This method has the added
benefit that the user does not need to model each individual grain in the material, which would lead to an increased computational burden. While a Monte-Carlo benchmark of the
method was shown in the previous study \cite{Yamamoto2014}, no experimental investigation of the method was presented. \\
\indent In the following, we first describe an experimental investigation of the particle self-shielding effect in our developed concrete using low-energy neutrons. We then give a description of the concrete, followed by a discussion
of the theoretical model and simulation methodology used to analyze the experimental data. Lastly, we present a comparison of the simulations and measured results. \\ 
\section{Experimental procedure}
The low-energy neutron measurements were carried out at the R2D2 test beamline at the JEEP II reactor at the Institute for Energy Technology located in Kjeller, Norway \cite{IFE}. Neutrons from the reactor core were incident
on a Ge wafer monochromator, of total crystal height 108 mm and width 54 mm, which reflected and focused neutrons of a given wavelength at a 90$^{o}$ take-off angle into a borated PE collimator. The 400 reflection of the Ge wafers was used to provide
neutrons of 2 {\AA}. Outside of the collimator, a neutron beam monitor provided an indication of the number of reflected neutrons exiting the collimator and two borated aluminum slits from JJ-XRAY \cite{jjxray} were used to define the divergence of the beam. The first slit had a opening width of 11.7 mm and a height of 43 mm while the second slit had an opening width of 7.6 mm and a height of 26 mm.
This allowed a maximum beam divergence of $\sim$3.6$^{o}$ in the vertical direction and $\sim$2.3$^{o}$ in the horizontal direction to pass through the slits. The concrete samples were placed after the slits and a $^3$He proportional counter \cite{GE} was placed some distance behind the position of the samples. The counter was placed in a borated PE housing, with an opening of width 5 mm and height 65 mm. Furthermore, a 2 cm thick B$_{4}$C slit with an opening width of 5 mm and height 25 mm was used to define the beam at the counter position.  An overview of the experimental setup outside of the collimator is shown in Fig. 1 and a photo is shown in Fig. 2. \\
\indent A detailed description of the PE-B4C-concrete is given previously in \cite{DiJulio2017}, however a brief overview is given here. The concrete was created by adding 10 wt\% of PE in the form of a 50-50 mix of 2.5 mm and 5.0 mm
diameter beads and a total of 0.76 wt\% of B$_{4}$C grains to a standard concrete mixture. This is equivalent to 20 vol\% PE and 0.6 vol\% B$_{4}$C, where the PE replaced the same volume of granite in the concrete and
the B$_{4}$C replaced the same volume of SiO$_2$. Due to these modifications, the specialized concrete had a lower density and lower compression strength than a standard concrete \cite{DiJulio2017}. The exact contents of the mixture are given in \cite{DiJulio2017} and the calculated elemental
content from this mixture is given in Table 1. \\
\indent Five different batches of the concrete were produced with different B$_{4}$C
grain sizes, however the total weight fraction of B$_{4}$C was kept the same. The grain size diameter groupings ranged from: less than 0.063 mm, 0.063 mm - 0.125 mm, 0.125 mm - 0.25 mm, 0.25 mm - 0.5 mm, and 0.5 mm - 1.0 mm.
Each group was created from a powder containing a random distribution of B$_{4}$C grains with the use of a sieve, thus the exact diameter distribution of the grains within each group is not known. Several different thicknesses of
concrete blocks were produced for each batch with total thicknesses of 2.5 cm, 5.0 cm, 7.5 cm and 10 cm. The samples were made at the Danish Technological Institute \cite{DTI} in small batches, with a total weight of 35 kg each. \\
\indent The samples were placed at the experimental setup as indicated in Fig. 1 and exposed to the neutron beam for a given time. The number of neutrons detected in the counter were recorded and saved for further analysis.
For the thicknesses of 2.5 cm and 5.0 cm, six measurements at different locations on each concrete block were performed. For the
the 7.5 cm thick samples, four such measurements were carried out while only one measurement was performed for the 10 cm thick samples, due to limited beam time. In the cases where more than one measurement was performed,
the average of the number of detected neutrons was calculated and used in the analysis below. The background in the detector was determined by tilting the monochromator off the 400 reflection peak while
leaving the concrete sample in place. Thus the background measurements included slowing down effects of higher-energy neutrons in the beam to lower energies and also contributions from other ambient sources
in the reactor hall. Background subtraction and normalization of the results were calculated using the total counting times of the individual measurements and the counts provided by the beam monitor.
The statistical uncertainty in the measurements was calculated using standard error calculation procedures. \\
\section{Simulation Methodology}
\subsection{Model Description}
The general theory behind the effective cross-section method is described in general in \cite{Shmakov2000,Yamamoto2006,Yamamoto2010} and \cite{Yamamoto2014} when applied to shielding calculations. An overview of the relevant aspects
of the theory is presented here for completeness, however an interested reader should see the indicated references for more specific details. \\
\indent If a slab of material contains a uniform random distribution of grains of a given diameter, $D$, it can be shown that the effective homogenized macroscopic total cross-section of the slab is given by
\begin{equation}
  \overline{\Sigma} = \Sigma_{m} - \frac{1}{L}ln[1-q+qJ(-{\triangle}{\Sigma},D)], \\
\end{equation}
where ${\triangle}{\Sigma}=\Sigma_a-\Sigma_m$, $\Sigma_a$ is the macroscopic total cross-section of the absorber material, $\Sigma_m$ is the macroscopic total cross-section of the matrix material in which the grains are dispersed in, $q=1.5\alpha$ with $\alpha$ being the volume fraction of the grains. The model assumes that the total thickness of the material is divided into sub-layers of thickness $L$. In this work, $L$ is taken to be equal to $D$, see the above mentioned references for more detail.
The function $J(x,y)$ is given by,
\begin{equation}
  J(x,y)=\frac{2}{y^2}\left(\frac{1}{x^2}+\left(\frac{y}{x}-\frac{1}{x^2}\right)exp(xy)\right).
\end{equation}
The effective microscopic cross-section of an isotope, $\sigma_e^{i,x}$, in the material can be calculated from $\sigma_e^{i,x}=F_{a,m,e}\cdot{}\sigma_{}^{i,x}$ where $F_{a,m,e}$ is the correction factor, $i$ and $x$ indicate an isotope and reaction type, and
$\sigma_{}^{i,x}$ is the tabulated microscopic cross-section of the isotope. The correction factors depend on whether the isotope exists in the absorber material, matrix material, or both. The correction factor for an isotope in the absober material is
\begin{equation}
  F_a=\frac{\overline{\Sigma}P_a}{\alpha\Sigma_aP_T},
\end{equation}
and for an isotope in the matrix material,
\begin{equation}
  F_m=\frac{\overline{\Sigma}(P_T-P_a)}{(1-\alpha)\Sigma_mP_T},
\end{equation}
and if the isotope exists in both the absorber grain and the matrix material, the correction factor is
\begin{equation}
  F_e=\frac{(1-\alpha)N_m^iF_m+\alpha{}N_a^iF_a}{(1-\alpha)N_m^i+\alpha{}N_a^i},
\end{equation}
where $N^i_{m,a}$ represents the atom density of the isotope in either the matrix material or the absorber grain. The parameters $P_a$ and $P_T$ are the collision probabilities in the absorber grain and matrix material, and are given by,
\begin{equation}
P_a=q\cdot{}exp(-\Sigma_mD/2)\cdot(J(\Sigma_m/2,D)-J(\Sigma_m/2-\Sigma_a,D)),
\end{equation}
and
\begin{equation}
P_T=1-exp(\overline{\Sigma}L).
\end{equation}
\subsection{Geant4 simulations}
Based on the theoretical description in the previous section, we have implemented a patch in Geant4 \cite{Geant4,Geant4Ref1} in order to calculate the cross-section correction factors (Eqs. (3-5)) during the simulation process.
The correction factors were implemented in the G4ParticleHP module and calculated at any time a neutron cross-section was accessed. The implementation takes into account the grain size effect due only to the
B$_{4}$C grains and not inhomogeneities related to the the PE or other concrete mixture ingredients. The version of Geant4 used was 10.3 patch-02. \\
\indent A model of the experimental setup was created in the standard way in Geant4 and included the description provided in Fig. 1. In order to compare directly with the measured results, the detector efficiency \cite{Piscitelli2014} was also included in the analysis. This was calculated using tools provided by the ESS detector group framework \cite{Kittelmann2014}. 
\section{Results and Discussion}
The results of the measurements are presented in table 2 and are compared with the Geant4 simulations in Fig. 3, where the quantity R$/$R$_0$ is plotted as a function of concrete total thickness. The quantity R is the measured or simulated response of the proportional counter
when the indicated concrete thickness was placed in the neutron beam while R$_0$ is the response when no blocks were placed in the beam. The dashed lines represent the results of the simulations for the upper and lower limits of the grain size distribution
for a given batch of concrete. The solid line represents the simulations if the effective cross-section method was not used, which assumed a uniform random distribution of B in the concrete.
The statistical uncertainties of the simulation results were no more than 6$\%$. The missing data point for the 10 cm thick block with 0.125-0.250 mm grain sizes was due to limited beam time.
The measurement using the 10 cm thick block with $<$0.063 mm grain sizes was close to the measured background level and thus the error analysis yielded a large uncertainty. \\
\indent In all cases shown in Fig. 3, the experimental results lie above the simulation results when not using the effective cross-section method. This shows that assuming a completely homogeneous distribution of B$_4$C without invoking the effective cross-section method
would lead to an over-estimation of the neutron absorption in the concrete. When the method was used, it can be seen that there is much improved agreement between the simulations and measurements. The small differences between the simulations and measurements
may be due to the fact that the concretes were created in small batches, as mentioned in section 2. Due to this, there could have been regions in the concrete which were not completely homogeneous and/or could contained air pockets. \\
\indent These results show that calculations not using the effective cross-section approach would result in estimates of thinner shielding than actually required for a specific design scenario. As the grain sizes get smaller, both the simulation and measured
results approach the simulation results without the effective cross-section method. This indicates that using as small as possible grain sizes would yield the best possible performance of the concrete. However, the choice of grain size also depends on the price of the
B$_{4}$C grains, which increases as the grain size decreases, and also on the stability of the concrete, which can decrease as the grain size gets smaller. 
\section{Conclusions}
In summary, we have carried out a combined measurement and simulation study on the low-energy neutron transport properties of a specialized neutron shielding concrete containing small B$_4$C grains. Five batches of the concrete were produced with varying grain sizes and tested
using a 2 {\AA} neutron beam at the R2D2 beamline at the Institute for Energy Technology in Kjeller, Norway. It was found that the concretes exhibited a reduced performance due to the particle self-shielding effect of the B$_4$C grains which could be well reproduced using an effective cross-section method
implemented in Geant4 simulations. This study validates the use of this method for simulating shielding materials based on small absorber grains for low-energy neutrons. \\

\section{Acknowledgments}
\noindent This project has received funding from the European Union’s Horizon 2020 research and innovation programme under grant agreement No 654000.
We would like to thank Rodion Kolevatov for useful discussions regarding various aspects of this work.

\section*{References}
\bibliographystyle{elsarticle-num}
\bibliography{ife_paper}

\begin{table}[h]
\caption{Material properties for the PE-B4C-concrete that was studied.}
\begin{center}
  \begin{tabular}{|c|c|}
    \hline
    Element & Weight percent \\
    \hline
O   & 46.06\% \\
Ca  & 8.05\% \\
Si  & 28.4\% \\
Al  & 2.34\% \\
Fe  & 0.837\%  \\
Mg  & 0.195\%  \\
Na  & 0.613\%  \\
K   & 1.25\% \\
S   & 0.276\% \\
Cl  & 0.00353\% \\
H   & 2.362\%\\
Ti  & 0.0517\%\\
P   & 0.0259\%\\
C   & 8.93\%\\
B   & 0.596\% \\
\hline
Density & 1.97 (g/cm$^{3}$) \\ 
\hline
\end{tabular}
\end{center}
\label{default}
\end{table}%

\begin{table}[h]
  \caption{The Measured results, R$/$R$_0$, as a function of concrete block thickness and B$_{4}$C grain size. The quantity R is the response of the proportional counter
when the indicated concrete thickness was placed in the neutron beam while R$_0$ is the response when no blocks were placed in the beam.}
\begin{center}
  \small
  \begin{tabular}{|l|l|l|l|l|l|}
    \hline
    Thickness & 0.500-1.000 mm & 0.250-0.500 mm & 0.125-0.250 mm & 0.063-0.125 mm &  $<$0.063 mm \\
    \hline
2.5 cm  & (5.14$\pm$0.87)$\times$10$^{-2}$ & (3.70$\pm$0.79)$\times$10$^{-2}$ & (2.58$\pm$0.70)$\times$10$^{-2}$& (3.08$\pm$0.86)$\times$10$^{-2}$ & (2.24$\pm$0.57)$\times$10$^{-2}$ \\
5.0 cm & (1.91$\pm$0.50)$\times$10$^{-3}$  & (1.06$\pm$0.19)$\times$10$^{-3}$ & (2.22$\pm$1.23)$\times$10$^{-3}$& (1.77$\pm$0.46)$\times$10$^{-3}$ & (4.90$\pm$1.65)$\times$10$^{-4}$\\
7.5 cm & (6.54$\pm$2.02)$\times$10$^{-5}$  & (5.85$\pm$1.38)$\times$10$^{-5}$ & (7.27$\pm$3.44)$\times$10$^{-5}$& (2.72$\pm$0.94)$\times$10$^{-5}$ & (3.92$\pm$1.26)$\times$10$^{-5}$\\
10.0 cm & (2.40$\pm$0.18)$\times$10$^{-5}$ & (7.96$\pm$1.73)$\times$10$^{-6}$ &                -               & (4.64$\pm$1.78)$\times$10$^{-6}$ & (0.13$\pm$1.73)$\times$10$^{-6}$\\
\hline
\end{tabular}
\end{center}
\label{default}
\end{table}%

\cleardoublepage

\begin{figure}[!t]
\centering
\includegraphics[width=140mm]{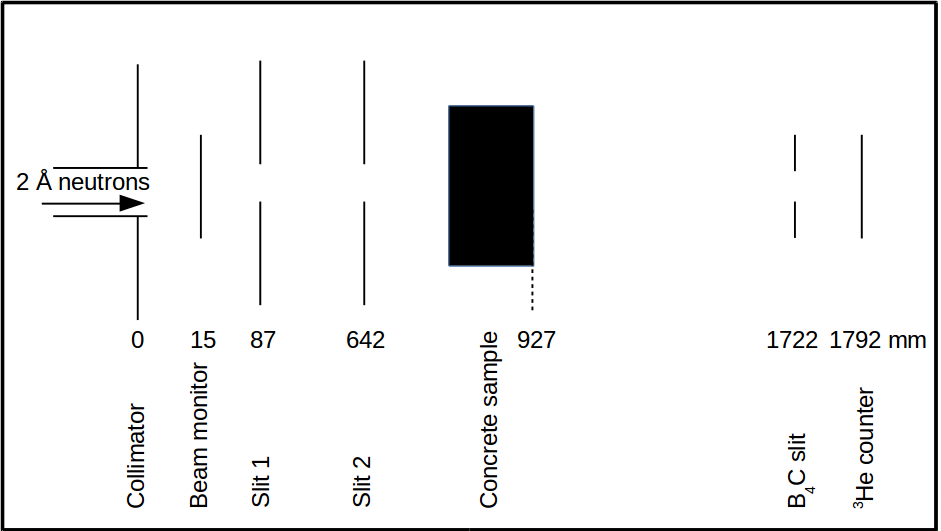}
\caption{A schematic of the setup used for the concrete measurements at the R2D2 test beamline.}
\label{fig:flowchart}
\end{figure}

\begin{figure}[!t]
\centering
\includegraphics[width=100mm]{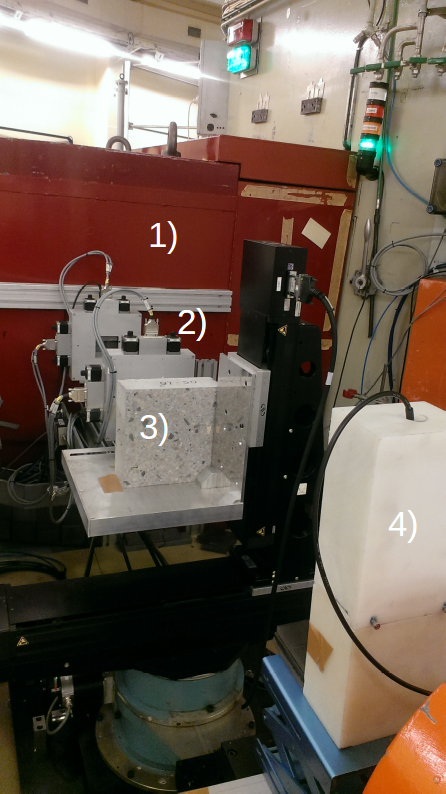}
\caption{A photo of the setup used for the concrete measurements at the R2D2 test beamline. The components indicated are 1) the monochromator shielding, 2) the slit system, 3) a concrete sample, and 4) the proportional counter and shielding.}
\label{fig:flowchart}
\end{figure}

\begin{figure}[]
  \vspace*{-1.8in}
  \hspace*{-0.5in}
\includegraphics[width=140mm]{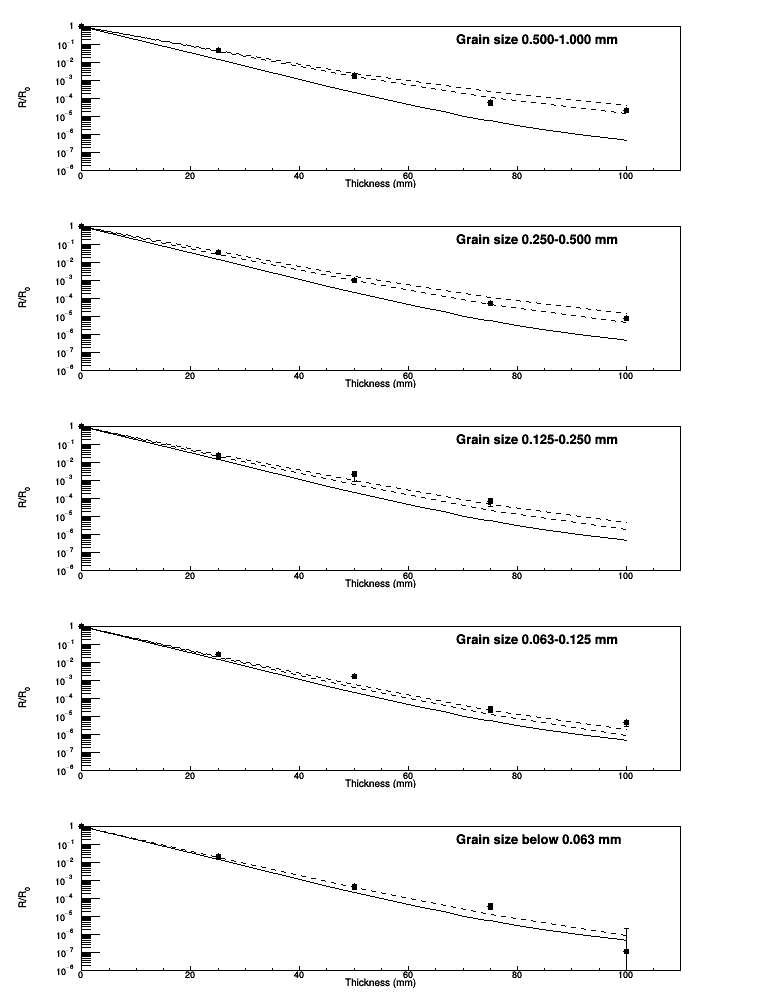}
\caption{Comparison between the measured results and the Geant4 simulation results. The dashed lines represent the ranges of the grain size boundaries used in the simulations, indicated by the text in each sub-figure. The solid lines represent
  the results of the Geant4 simulations if a homogeneous mixture of B$_4$C is assumed without the effective cross-section method. The points indicate the results of the measurements.
The quantity R is the measured or simulated response of the proportional counter when the indicated concrete thickness was placed in the neutron beam while R$_0$ is the response when no blocks were placed in the beam.}
\end{figure}

\end{document}